\documentclass[a4paper,11pt]{article}
\usepackage{latexsym}	
\usepackage{graphicx}
\usepackage{amsmath}
\usepackage{amsthm}

\usepackage{ amssymb }
\def\ba{\begin{eqnarray}}
\def\ea{\end{eqnarray}}

\def\ba{\begin{eqnarray}}
\def\ea{\end{eqnarray}}

\def\lb{\label}
\def\be{\begin{equation}}
\def\ee{\end{equation}}

\theoremstyle{plain}

\begin{document}
\title{Neutrino emission and initial evolution of axionic quark nuggets}
 \author{ Osvaldo P. Santill\'an \thanks{Instituto de Matem\'atica Luis Santal\'o (IMAS), UBA CONICET, Buenos Aires, Argentina
firenzecita@hotmail.com and osantil@dm.uba.ar.} and Alejandro Morano \thanks{Departamento de F\'isica, UBA, Buenos Aires, Argentina}}

\date {}
\maketitle

\begin{abstract}

The axion quark nuggets introduced in \cite{zhitnitsky}-\cite{zhitnitsky13} are a candidate for cold dark matter which, in addition, may be relevant in baryogenesis scenarios. 
The present work studies the evolution of these objects till they enter in the colour superconducting phase. This evolution was already considered
in \cite{zhitnitsky5}, where it is concluded that a large chemical potential $\mu$ is induced on the bulk of the object. That work takes the baryon number accumulated at the domain wall surrounding the object as predominant, and suggests that internal and external fluxes are compensated in such a way that they not modify considerably the dynamics of the object if they are neglected. In the present work the possibility that the bulk contribution to the baryon number may be relevant at initial stages, and  that the object may emit a large amount of neutrinos due to quark-antiquark annihilations is taken into account.
This results into a more violent contraction of the object and perhaps a more effective cooling. The outcome is that the formed objects may have an smaller size. Even taking into account these corrections, it is concluded that the
cosmological applications of these objects  are not spoiled. These applications are discussed along the text.

\end{abstract}

\section{Introduction}

One  important problem in cosmology and particle physics is to understand if the present universe is baryon asymmetric or if the anti-baryons are segregated
from baryons on very large scales. If there were galaxies of matter and anti matter in a given cluster of galaxies then, due to the presence of inter cluster gases, there should be nucleon anti-nucleon annihilations
leading to strong $\gamma$ ray emissions \cite{kolb}. As this effect is not observed, and since galaxies like Virgo contains around $10^{14}\text{M}_{\odot}$ of matter, it is believed anti-matter should
be segregated from matter on scales larger than $10^{14}\text{M}_{\odot}$. On the other hand, if the universe was initially baryon symmetric,
then  nucleons and anti-nucleons will freeze out at a temperature value $T\sim 22$ MeV. The ratio between the baryon number and the entropy densities  that remains at this temperature is around  nine 
orders of magnitude smaller than the observed value.  This discrepancy may be avoided if there is an unknown segregation mechanism between baryons and anti-baryons which takes place at $T\sim 41$ MeV \cite{kolb}. However, the Hubble horizon
at these temperature is considerably smaller than $10^{14}\text{M}_{\odot}$.
A possible solution is that the universe at $T\geq 41$ MeV was already in a baryon asymmetric state. There are particle physics scenarios which predict  non zero baryon number \cite{kolb}, several of them are based
on the Sakharov requirements for baryogenesis \cite{sakharov}-\cite{sakharov2}. These requirements include in particular C and CP violation, and tiny baryon number violating interactions at the beginning of the universe.

Another possible explanation for baryogenesis was presented in \cite{zhitnitsky}-\cite{zhitnitsky13}. This scenario is based on an apparently 
unrelated problem namely, the axion solution of the CP problem in QCD \cite{axion1}-\cite{sikivie}. The axion is a pseudo scalar particle $a$ that has several cosmological applications related to the formation of topological defects. An example
are domain walls. At first sight, it is believed that such walls are problematic, as their evolution should overcome the critical energy density $\rho_c\sim 8.27\; 10^{-27}$kg/m$^3$ which will lead to a catastrophe with observations, as discussed in \cite{sikivie}
and references therein. But there exist the so called $N=1$ axion models
for which this problem does not exist, and there exist several ideas on how solving this problem for other types of axion models \cite{sikivie}. The authors of \cite{zhitnitsky}-\cite{zhitnitsky13} assume that at a temperature $T\sim 100$ MeV
there exists a network of domain walls admitting non trivial quark configurations which, due non trivial asymptotic conditions, carry non zero baryon number \cite{zhitnitsky5}.
Owing to the domain wall tension $\sigma$, some of these regions tends to contract and press the quark-gluon plasma trapped inside. As a consequence, the system acquires a non zero chemical potential $\mu$ and contracts until the internal Fermi pressure equals the external one, and then realises damped oscillations around the equilibrium radius $R_e$. The resulting internal temperature $T_i$ is small enough, and the chemical potential $\mu$ is large enough, for reaching the colour superconducting (CS) phase in the bulk of the object \cite{cfl}-\cite{cfl2}.  It is important to emphasize that binding energy  $\Delta$ characteristic of this state is large enough for these  objects  not participating in nucleosynthesis at  $T\sim 1$ MeV \cite{zhitnitsky2}.

The formation of the axion nuggets described above however, does not explain by itself the baryon asymmetry of the universe. Despite these objects contain net baryon or anti-baryon number, it is expected the presence of equal quantity of them if the underlying physics does contain baryon number violating processes. Thus, a further mechanism for asymmetry generation should be found. A crucial point for generating a larger number of anti-baryon objects may be the dynamics of the coherent axion field, which may lead to a preferential evolution in favor of anti nuggets. This mechanism is effective regardless
the small value of the $\theta$ term as long as it remains coherent on the universe scale during the formation process \cite{zhitnitsky5}. This is a new feature, not present in the ordinary quark nuggets models such as \cite{bodmer}-\cite{witten}.
These hypothesis are reasonable from the physical point of view. However, a precise quantitative analysis about the resulting asymmetry is technically involved and, at the moment, is lacking. There are other type of segregation mechanisms such as \cite{atreya} for baryogenesis, but these will not be considered in the present paper.

There are some special features that distinguish axion quark nuggets from other dark matter candidates. First, these axion lumps are not supposed to be weakly interacting with ordinary matter. Their interaction is strong in fact, but they are macroscopically large as well.
For this reason, the quotient between the cross section for interaction with visible baryons to their mass $\sigma/M\sim 10^{-10}$cm$^2$/g, which is well below  the typical astrophysical limits $\sigma/M\sim $ cm$^2$/g \cite{zhitnitsky2}.  Another salient characteristic is that these compact objects are long lived, with mean life time larger than the present age of the universe. In fact, it has been suggested that the excess of $\gamma$
ray flux in MeV and GeV bands may be explained in terms of the rare annihilations between these objects and ordinary baryons. In addition, these objects interact noticeably with photons. However, the mean free time of photons 
when colliding with these objects is much larger than the Hubble time, thus these objects can be considered as cold dark matter, even when they are not electromagnetically neutral. These characteristics 
makes these objects different, for instance, than WIMPs. Further details are discussed in \cite{zhitnitsky2}.

The evolution of these objects till they enter in the CS phase \cite{cfl}-\cite{cfl2} was  considered in \cite{zhitnitsky5}. In the present work, a variant  for the dynamics of such objects is described. Although there are some different details about the nature of the formation in comparison with those pointed \cite{zhitnitsky5}, the main 
cosmological applications remain valid. The main differences are the following. In studying the fate of the axion nuggets, the authors of \cite{zhitnitsky5} reach to the conclusion 
that a large chemical potential $\mu$ is induced at the object bulk. These authors assume that main contribution to the baryon number of the object
is given by the wall contribution, which carry non zero baryon number even in the limit of $\mu\to 0$, and neglect the effect of internal and external fluxes. In the present work it is assumed that, besides this wall contribution, a considerable bulk baryon number may appear at early stages of the evolution. In addition, it is assumed that there is a considerable neutrino emission due to quark anti-quark annihilations. 
The neutrino emissivity plays the role of expulsion of fuel, and generates a violent contraction of the object. Even taking into account
these circumstances, we are able to estimate that the object is formed when the universe temperature is around $T\sim 41$ MeV and that it falls into the CS phase. The only difference is the size of the object, but it will be argued in the text that this does not spoil the main cosmological applications of these nuggets.

 It should be emphasized that there are effects that are neglected, for instance neutrino or other particle adsorbtion. But even taking into account that we are employing the most unfriendly conditions and still being able of obtain plausible results, the present scenario gives a hint that axion quark nuggets may be a candidate for both cold dark matter production and baryogenesis.

The present work is organised as follows. In section 2, the general form of the equations of motion for these objects is described. In section 3 these equations are expressed through thermodynamical quantities such as internal temperature, chemical potential and the radius of the object. In section 4 the neutrino emission is estimated, by assuming that initially quark anti-quark annihilations play the leading important role at the initial evolution. In section 5 the fate of the nugget till it enters in the colour superconducting phase is described.  Section 6 contains the discussions of the obtained results.

\section{The generic equations of motion}

The initial state of an axion quark nugget is an axion domain wall enclosing some finite volume region \cite{zhitnitsky}-\cite{zhitnitsky13}. The exterior and interior are assumed 
to be in a quark-gluon plasma state, both with zero chemical potential \cite{zhitnitsky5}. The external region may fall into the hadron phase at some point during the evolution of the object, however, this will not affect significantly the following description.
 The initial temperature of the universe is approximately at $T_0\sim 100-150$ MeV which, to the standard history of the universe, corresponds to a time $t\sim 10^{-4}$s. The domain wall tension $\sigma$
 tends to contract the object, until the internal Fermi pressure equals the surface tension, and the wall then realises damped oscillations around the equilibrium position. There are quark degrees 
 of freedom living on the wall, therefore these objects carry non trivial baryon number even when their chemical potential $\mu$ vanishes inside and outside. 
This can be briefly explained as follows.  The equations of motion of a Dirac fermion $\Psi$ on a domain wall solution can be derived from the lagrangian \cite{zhitnitsky5}
\be\lb{dirac}
{\cal L}=i\overline{\Psi}[\gamma^\mu \partial_\mu-m e^{i[\theta(z)-\phi(z)]\gamma_5}-\mu\gamma_0]\Psi.
\ee
The fields $\theta(z)$ and $\phi(z)$ describe the axion and $\eta'$ fields constituting the wall. Here the four dimensional problem has been reduced to a two dimensional one,
with $\hat{z}$ a unit vector normal to the surface of the wall. By neglecting the back reaction of the fermions on the domain wall, there
exist non trivial fermion degrees of freedom $\Psi(z)$ that can live in the wall. These non trivial solutions carry a non zero baryon number
\be\lb{barnum}
N=\int \overline{\Psi}\gamma_0\Psi dx^3,
\ee
a number that is non vanishing due to non trivial asymptotic \cite{zhitnitsky5}.  

The value of the QCD axion constant is believed to be in the range $10^{9}$GeV$<f_a<10^{12}$GeV \cite{lower}-\cite{upper}. The value  to be employed here is close to $f_a\sim 10^{12}$GeV, which means that the axion mass is close to the value
 \be\lb{axionmass}
 m_a\sim \frac{m_\pi f_\pi}{f_a}\sim 10^{-5} \text{eV}.
  \ee
This complements in some sense the results of \cite{zhitnitsky4}, as this reference considers a value close to $f_a\sim 10^{10}$GeV. Neglecting the fermions backreaction, the surface tension acting inwards on the bubble is given by \be\lb{sigma}\sigma\sim 8 f_a m_\pi f_\pi\sim 10^{20}\text{MeV}^3,\ee for 
 the choice of $f_a$ given above. An important point to be discussed is the initial size of the nuggets. The argument of \cite{zhitnitsky4} is based on the Kibble mechanism \cite{kibble}-\cite{kibble2}. In this scenario, the early universe at a temperature $T_0\sim 100-150$ MeV is composed by a percolated cluster of domain walls of very complicated topology \cite{vilenkin}. There are numerical simulations reviewed in the book \cite{vilenkin} that suggest that approximately the 0.87 of the total universe wall area belongs to the percolated cluster, while the remaining part is represented by small closed bubbles. This small fraction is crucial for the purposes of \cite{zhitnitsky4}-\cite{zhitnitsky5}, as is enough for the axionic quark nuggets to form \cite{zhitnitsky5}. This portion of domain wall energy does not contribute considerably to axion production considered in \cite{sikivie}, \cite{sikivie2}-\cite{sikivie3}.
Concerning the correlation length $\xi$, one possibility is to take  $\xi\sim m_a^{-1}\sim 1$ cm as a characteristic initial scale, which is the type of length considered in \cite{zhitnitsky4}. The probability of finding closed domain walls of $R>>\xi$ is exponentially suppressed \cite{kibble}-\cite{kibble2}

\footnote{As is clear from the above discussion, the reference \cite{zhitnitsky4} involves only closed bubbles with no strings attached. The present paper follows this approach. But there exist scenarios which do not rely in the Kibble mechanism at a QCD scale, examples are \cite{sikivie2}-\cite{sikivie3}. The initial size and density of such defects will be different for these models. Examples with such characteristics and their possible phenomenological consequences will be discussed at the end of this work, see the last section.}. 
 
The purpose of the present work is to understand qualitatively the fate of this region as it contracts.
The equations of motion of the bubbles just described will be taken schematically as follows
\be\lb{newton}
\frac{d P^\alpha_R}{d\tau}+\frac{d P^\alpha_\nu}{d\tau}=F^\alpha,
\ee
with $\alpha=0,1,2,3$ and $\tau$ the proper time of the event.  Equations of this type describe the motion of a relativistic rocket whose mass $M$ varies with time due to the expulsion of fuel.
The role of the fuel is played by the loss of neutrinos, whose momentum was denoted above by $P^\alpha_\nu$. In the last expression, $P^\alpha_R$ denotes the 4-momentum of an infinitesimal mass element $dM$ composing the bubble.
In addition, the 4-force acting on the system  $$F^\alpha=\frac{1}{\sqrt{1-\dot{R}^2}}(f\cdot \dot{R}, f^i),$$ has been introduced.
Here $f^i$ being the force applied over the system, which is radially directed in the frame located at the center of the bubble.
In practice, this force will be the sum of the surface tension force, and the one arising from the internal and external pressures. The bubble itself is considered as the sum of all these infinitesimal elements,
simultaneously moving in the radial direction. The neutrinos are  assumed to be emitted isotropically.  

The equations written above are covariant, that is, they are valid in any inertial frame. The derivatives with respect to the proper time $\tau$ of  the 4-vectors  $P^\alpha$  in (\ref{newton}) can be related to a coordinate time $t$ of the reference system located at the center of the bubble by
$$
\frac{d}{d\tau}=\frac{1}{\sqrt{1-\dot{R}^2}}\frac{d}{dt}.
$$
The momentum of a surface element of the wall with respect to this reference system is
$$
P_R^0=\frac{dM}{\sqrt{1-\dot{R}^2}},\qquad P_R^i=\frac{\dot{R}dM}{\sqrt{1-\dot{R}^2}}.
$$
The neutrino momentum is such that $(P^0_\nu)^2=P_\nu^i P_\nu^i$ since it can be considered as a relativistic particle.
The transformation of the momentum $P_\nu$ from frame of the wall to the momentum from center of the bubble $P'_\mu$ is given by
$$
P_\nu^{'\alpha}=\sqrt{\frac{1-\dot{R}}{1+\dot{R}}}P_\nu^\alpha.
$$
On the other hand
$$
\frac{dP_\nu^{'\alpha}}{d\tau}=\frac{dP_\nu^{'\alpha}}{dP_\nu^\beta}\frac{dP_\nu^\beta}{d\tau}.
$$
By combining the last two formulas it is obtained that
$$
\frac{dP_\nu^{'\alpha}}{d\tau}=\sqrt{\frac{1-\dot{R}}{1+\dot{R}}}\frac{dP_\nu^\beta}{d\tau}.
$$
By integrating along the solid angle $d\Omega$ and by simplifying a common $\sqrt{1-\dot{R}^2}$ factor, the equation (\ref{newton}) can be expressed as follows
\begin{equation}\label{ecomoco2}
\frac{d}{dt}\bigg(\frac{M}{\sqrt{1-\dot{R}^2}}\bigg)+\sqrt{\frac{1-\dot{R}}{1+\dot{R}}}\frac{dP_\nu}{dt}=4\pi R^2\dot{R}\Delta P,
\end{equation}
\begin{equation}\label{ecomoco}
\frac{d}{dt}\bigg(\frac{M\dot{R}}{\sqrt{1-\dot{R}^2}}\bigg)+\sqrt{\frac{1-\dot{R}}{1+\dot{R}}}\frac{dP_\nu}{dt}=4\pi R^2\Delta P.
\end{equation}
These equations represent a variable mass nugget emitting neutrinos as fuel, and acted by radial forces.
They will be supplemented below, when applies, with a further constraint arising from the conservation of the baryonic number of the system.

\section{The explicit equations of movement}

The emission of neutrinos is usually described in terms of the so called emissivity $Q_\nu$ \cite{emissivity1}-\cite{emissivity12} through the relation
\be\lb{emisivitys}
\frac{dP_\nu}{dt}=\frac{4\pi R^3}{3}Q_\nu.
\ee
The emissivity $Q_\nu$ will be characterised in the next section. But at this point, it may be convenient to describe in detail the other quantities appearing in the equations (\ref{ecomoco2})-(\ref{ecomoco}). This system  can be rewritten as follows
\begin{equation}\label{ecomocos2}
\frac{d}{dt}\bigg(\frac{M}{\sqrt{1-\dot{R}^2}}\bigg)+\sqrt{\frac{1-\dot{R}}{1+\dot{R}}}\frac{4\pi}{3} R^3 Q_\nu=4\pi R^2\dot{R}\Delta P,
\end{equation}
\begin{equation}\label{ecomocos}
\frac{M\ddot{R}}{\sqrt{1-\dot{R}^2}}+\bigg[4\pi R^2\dot{R}\Delta P-\sqrt{\frac{1-\dot{R}}{1+\dot{R}}}\frac{4\pi}{3} R^3 Q_\nu\bigg]\dot{R}+\sqrt{\frac{1-\dot{R}}{1+\dot{R}}}\frac{4\pi}{3} R^3Q_\nu=4\pi R^2\Delta P.
\end{equation}
Note that the equation (\ref{ecomocos2}) is the same as (\ref{ecomoco2}). On the other hand, the equation (\ref{ecomocos}) is obtained by (\ref{ecomoco}) by replacing the derivative of $M(1-\dot{R}^2)^{-\frac{1}{2}}$ with respect to $t$
through (\ref{ecomoco2}). Now, if the motion of the bubble is non relativistic, that is $\dot{R}<<1$, then these equations may be reduced to
\begin{equation}\label{ecomocoso2}
\frac{dM}{dt}+\frac{4\pi}{3} R^3 Q_\nu=4\pi R^2\dot{R}\Delta P,
\end{equation}
\begin{equation}\label{ecomocoso}
M\ddot{R}+\frac{4\pi}{3} R^3Q_\nu=4\pi R^2\Delta P.
\end{equation}
In order to solve these equations, the mass $M$ and the pressure forces acting on the bubble should be characterised. In the following, it will be assumed that, during short periods, the state of the bubble
may be approximated by an equilibrium state with well defined temperature $T$, chemical potential $\mu$ and internal pressure $P_i$.
 In this case, the mass $M$ of the bubble is given by \cite{zhitnitsky5}
\be\lb{masamasa}
M=E=4\pi R^2 \sigma+\frac{4\pi}{3} R^3\rho+\frac{4\pi}{3} R^3 E_B\Theta(\mu-\mu_1)\bigg[1-\frac{\mu_1^2}{\mu^2}\bigg],\qquad E_B\sim (150\; \text{MeV})^4.
\ee
Here $E_B$ is the bag constant, which should be taken into account when the chemical potential $\mu$ is higher than $\mu_1\sim 330$ MeV \cite{zhitnitsky4}. The surface energy is sourced by the surface tension $\sigma\sim8  f_a m_\pi f_\pi$ of the axion wall.
On the other hand, for generic fermions with mass $m$ the pressure is related to the energy density $\rho$ as follows
$$
P=\frac{\rho}{3}-\frac{m^2g}{6\pi^2}\int_m^\infty \frac{\sqrt{\epsilon^2-m^2}d\epsilon}{e^{\frac{\epsilon-\mu}{T}}+1},\qquad \rho=\frac{g}{2\pi^2}\int_m^\infty \frac{\epsilon^2\sqrt{\epsilon^2-m^2}d\epsilon}{e^{\frac{\epsilon-\mu}{T}}+1}.
$$
The last two formulas suggests that the pressure grows as $m$ decreases.  As the $u$ and $d$ quarks masses are of the order $m_u\sim m_d\sim 4$ MeV and the initial universe temperature is of the order $T_0\sim 100$ MeV, one may consider
these quarks as massless. The $s$ quark mass is $m_s\sim T_0$ and the pressure contribution of this species is smaller than the lighter counterparts. For this reason,  the simplifying assumption that quark are massless will be employed when calculating some thermodynamical properties, as it will not lead to a significant deviation of their real values.
Under this approximation, it follows that
$$
P_q=\frac{1}{3}\rho-E_B \Theta(\mu-\mu_1)\bigg[1-\frac{\mu_1^2}{\mu^2}\bigg],\qquad \rho=\rho_q+\rho_{\overline{q}}=\frac{g T^4}{2\pi^2}\bigg[\int_0^\infty \frac{x^3 dx}{e^{x-\beta}+1}+\int_0^\infty \frac{x^3 dx}{e^{x+\beta}+1}\bigg],\qquad \beta=\frac{\mu}{T},
$$
with explicit result given by
\be\lb{ro}
P_q=\frac{1}{3}\bigg\{\frac{7}{8}\frac{\pi^2 g T^4}{30}\bigg[1+\frac{30\beta^2}{7\pi^2}+\frac{15\beta^4}{7\pi^4}\bigg]-3E_B \Theta(\mu-\mu_1)\bigg[1-\frac{\mu_1^2}{\mu^2}\bigg]\bigg\}.
\ee
Here the effect of the bag constant $E_B$ has been included, which tends to decrease the pressure when it is turned on inside the compact object.
In all the formulas derived above, the degeneracy $g$ is 
$$
g=4 N_c N_f,
$$
since there are $2$ spin states and $2$ charge states (particle and antiparticle) for any flavour and colour. In addition, the number of colours is $N_c=3$.
The external pressure, or universe pressure, is given by \cite{zhitnitsky5}
\be\lb{expres}
P_e=\frac{7}{8}\frac{\pi^2 g T_e^4}{90},\qquad T_e=T_0\sqrt{\frac{t_0}{t}},\qquad T_0\sim 100-150\;\text{MeV},\qquad t_0\sim 10^{-4}\;\text{sec}.
\ee
The pressure difference acting on the surface of the bubble is then
\be\lb{pressurediff}
\Delta P=-\frac{2\sigma}{R}+\frac{7\pi^2 g T^4}{360}\bigg[1+\frac{30\beta^2}{7\pi^2}+\frac{15\beta^4}{7\pi^4}\bigg]-\frac{7\pi^2 g T_e^4}{360}-E_B \Theta(\mu-\mu_1)\bigg[1-\frac{\mu_1^2}{\mu^2}\bigg].
\ee
Note that in the last formula the effect of the surface tension $\sigma$ has been taken into account. In terms of the  thermodynamical expressions found above, the equations (\ref{ecomocoso2})-(\ref{ecomocoso}) can be expressed as follows
$$
\frac{d}{dt}\bigg[8\pi R^2 \sigma+\frac{2}{3}\frac{4\pi R^3}{3}\frac{7}{8}\frac{\pi^2 g T^4}{30}\bigg(1+\frac{30\mu^2}{7\pi^2T^2}+\frac{15\mu^4}{7\pi^4T^4}\bigg)+\frac{4\pi R^3}{3}\frac{7}{8}\frac{\pi^2 g T_e^4}{90}+\frac{8\pi R^3}{3}E_B \Theta(\mu-\mu_1)\bigg(1-\frac{\mu_1^2}{\mu^2}\bigg)\bigg]
$$
\be\lb{segun}
+\frac{4\pi R^3}{3}\frac{d}{dt}\bigg[\frac{7}{8}\frac{\pi^2 g T^4}{90}\bigg(1+\frac{30\mu^2}{7\pi^2T^2}+\frac{15\mu^4}{7\pi^4T^4}\bigg)+E_B \Theta(\mu-\mu_1)\bigg(1-\frac{\mu_1^2}{\mu^2}\bigg)\bigg]+\frac{4\pi}{3} R^3 Q_\nu=0,
\ee
$$
\bigg[4\pi \sigma R^2+\frac{4\pi R^3}{3}\frac{7}{8}\frac{\pi^2 g T^4}{30}\bigg(1+\frac{30\mu^2}{7\pi^2T^2}+\frac{15\mu^4}{7\pi^4T^4}\bigg)+\frac{4\pi R^3}{3}E_B \Theta(\mu-\mu_1)\bigg(1-\frac{\mu_1^2}{\mu^2}\bigg)\bigg]\ddot{R}+\frac{4\pi}{3} R^3 Q_\nu
$$
\be\lb{prime}
=-8\pi \sigma R +4\pi R^2\frac{7}{8}\frac{\pi^2 g T^4}{90}\bigg(1+\frac{30\mu^2}{7\pi^2T^2}+\frac{15\mu^4}{7\pi^4T^4}\bigg)-4\pi R^2E_B \Theta(\mu-\mu_1)\bigg(1-\frac{\mu_1^2}{\mu^2}\bigg)-4\pi R^2\frac{7}{8}\frac{\pi^2 g T_e^4}{90}+F_\eta.
\ee
Here an additional force on the bubble $F_\eta$ has been included which, at initial stages is not important. This is the QCD viscosity \cite{viscosity}
$$
F_\eta=\eta R \dot{R},
$$
which, for a contracting bubble, points outwards the surface of the bubble. This force is the result of several effects that occur during the 
contraction such as scattering of quarks, gluons and different Nambu-Goldstone bosons arising in different phases. The viscosity coefficient may depend on the temperature
and chemical potential $\eta(T, \mu)$. The value $\eta\sim m_\pi^3\sim 0.002$ GeV$^3$ will be employed in the following \cite{viscosity}. However, a further knowledge of the behaviour 
of $\eta$ as a function of $T$ and $\mu$ is of course desirable, especially in the limit $\mu>>T$. 

The equations (\ref{segun})-(\ref{prime}) constitute two equations for the three unknowns $T$, $\mu$ and $R$ as functions of the time parameter $t$.
The missing equation is related to the baryon number conservation of the system.
The baryon number of the system for these bubbles is initially localised on the axionic wall and its approximate expression is 
\be\lb{ba}
B=4\pi N g R^2\int \frac{ d^2p}{(2\pi)^2}\frac{1}{e^{\frac{\epsilon(p)-\mu}{T}}+1},\qquad \epsilon=\sqrt{p^2+m^2}.
\ee
Here $N$ is given by the expression (\ref{barnum}) and the wall was taken as a two dimensional object.
These integrals, in the massless limit, can be expressed in terms of the variable $\epsilon$ by taking into account that $p dp=\epsilon d\epsilon$. The result is
\be\lb{ecomocos33}
B=2 N g R^2T^2\bigg[\text{Li}_2(-e^{-\frac{\mu}{T}})+\frac{\pi^2}{6}+\frac{1}{2}\bigg(\frac{\mu}{T}\bigg)^2\bigg]. 
\ee
Since $\mu$ is positive, the argument $z$ of the dilogarithm function $\text{Li}_2(z)$ is such that  $|z|=e^{-\frac{\mu}{T}}<1$.
For this range of values, the dilogarithm may be expanded to give
\be\lb{ecomocos31}
B=2 N g R^2T^2\bigg[\frac{\pi^2}{6}+\frac{1}{2}\bigg(\frac{\mu}{T}\bigg)^2-\frac{\pi^2}{12}e^{-\frac{\mu}{T}}\bigg]. 
\ee
However, as the chemical potential $\mu$ grows, a volume contribution is turned on. The total baryon contribution
$$
B=2 N g R^2T^2\bigg[\frac{\pi^2}{6}+\frac{1}{2}\bigg(\frac{\mu}{T}\bigg)^2-\frac{\pi^2}{12}e^{-\frac{\mu}{T}}\bigg]+\frac{4\pi R^3}{3}(n_f-\overline{n}_f),
$$
where the last term is proportional to the difference between particles and anti-particles in a given volume. The variable $n_f$ is not independent with $\mu$ and $T$, in fact in the massless limit one has
$$
n_f-\overline{n}_f=\frac{gR^3 T^3}{6}\frac{\mu}{T}\bigg[1+\frac{1}{\pi^2}\bigg(\frac{\mu}{T}\bigg)^2\bigg].
$$
In this limit, the total baryon number becomes
\be\lb{ecomocos311}
B=2 N g R^2T^2\bigg[\frac{\pi^2}{6}+\frac{1}{2}\bigg(\frac{\mu}{T}\bigg)^2-\frac{\pi^2}{12}e^{-\frac{\mu}{T}}\bigg]+\frac{2\pi gR^3 T^3}{9}\frac{\mu}{T}\bigg[1+\frac{1}{\pi^2}\bigg(\frac{\mu}{T}\bigg)^2\bigg].
\ee
The baryon number conservation law is then expressed as
\be\lb{ecomocos3}
B=2 N g R^2T^2\bigg[\frac{\pi^2}{6}+\frac{1}{2}\bigg(\frac{\mu}{T}\bigg)^2-\frac{\pi^2}{12}e^{-\frac{\mu}{T}}\bigg]+\frac{2\pi gR^3 T^3}{9}\frac{\mu}{T}\bigg[1+\frac{1}{\pi^2}\bigg(\frac{\mu}{T}\bigg)^2\bigg]=\frac{Ng\pi^2}{3} R_0^2T_0^2,
\ee
where $T_0$ and $R_0$ are the quantities at the beginning of the formation, and it is assumed that $\mu_0=0$.
The equation (\ref{ecomocos3}) together with (\ref{segun})-(\ref{prime}) constitute a system of three equations determining the temperature $T$, the chemical potential $\mu$ and the radius $R$
of the object in terms of the initial conditions. This description is analogous of \cite{zhitnitsky5} but with the baryon volume term and neutrino emissivity $Q_\nu$ turned on.
In order to study the properties of their solutions, the expressions describing the emissivity $Q_\nu$ should be found. This will be done in the following section.

\section{The neutrino momentum release}

\subsection{General emissivity formulas}
As stated above, the neutrinos are assumed to be emitted isotropically due to 
pair annihilation in the bulk and at the border of the spherical region. The derivative of the momentum at a frame instantly at rest with respect to the domain wall is given in terms of 
the neutrino emissivity (\ref{emisivitys}). For the emissivity, there are several channels to consider and there is extensive literature about the subject, with possible applications to neutron stars \cite{emissivity1}-\cite{emissivity12}. However, for high temperatures $T>>\mu$ it will be assumed that
quark-antiquark annihilation in two neutrinos $q+\overline{q}\to \nu+\overline{\nu}$ is the leading channel.
The relevant coupling terms between the quarks and the neutrinos are given by
$$
{\cal L}_{q\nu}=\frac{G_F}{\sqrt{2}}\bigg[\overline{\nu}\gamma^\mu \frac{1+\gamma_5}{2}\nu\bigg]\bigg[\overline{u}\gamma^\mu (A_u+B_u\gamma_5)u+\overline{d}\gamma^\mu (A_d+B_d\gamma_5)d+\overline{s}\gamma^\mu (A_s+B_s\gamma_5)s\bigg],
$$
where the following parameters
$$
A_u=\frac{1}{2}-\frac{4}{3}\sin^2\theta_W,\qquad B_u=\frac{1}{2},
$$
\be\lb{pira}
A_d=-\frac{1}{2}+\frac{2}{3}\sin^2\theta_W,\qquad B_d=-\frac{1}{2},
\ee
have been introduced. The Weinberg angle $\theta_W$ is such that $\sin^2 \theta_W\sim 0.23$. We ignore the coupling for the $s$ quarks, but we assume that they are smaller than for the light quarks. A discussion about this coupling
will be given in the next section. By use of the above formulas the expression of the emissivity may be found,  which follows as a generalisation of the formula of electron emissivity \cite{emissivity12} adapted to quarks. For instance, the emissivity for a given quark $u$ is calculated by means of the following formula \cite{emissivity12}
$$
Q_u\sim \frac{G^2_F m_u^9}{36\pi}\bigg\{A_{u+}^2 \bigg[8(\Phi_{1u}U_{2u}+\Phi_{2u}U_{1u})-2(\Phi_{-1u}U_{2u}+\Phi_{2u}U_{-1u})+7(\Phi_{0u}U_{1u}+\Phi_{1u}U_{0u})
$$
\be\lb{pairemisivity}
+5(\Phi_{0u}U_{-1u}+\Phi_{-1u}U_{0u})\bigg]+9 A_{u-}^2\bigg[\Phi_{0u}(U_{1u}+U_{-1u})+U_{0u}(\Phi_{1u}+\Phi_{-1u})\bigg]\bigg\},
\ee
up to a factor related to colour matrices which is not far to unity.  In the last expression, the following thermodynamical integrals
\be\lb{termic}
U_{ku}=\frac{1}{\pi^2}\int_0^\infty \frac{p^2_u dp_u}{m_u^3}\bigg(\frac{\epsilon_u}{m_u}\bigg)^k f_u,\qquad \Phi_{ku}=\frac{1}{\pi^2}\int_0^\infty \frac{p^2_u dp_u}{m_u^3}\bigg(\frac{\epsilon_u}{m_u}\bigg)^k f_{\overline{u}},
\ee
and the following parameters 
$$
A^2_{+u}=A_u^2+B_u^2,\qquad A^2_{-u}=A_u^2-B_u^2,
$$
have been introduced. Here 
$$
f_u=\frac{1}{e^{\frac{\epsilon_u-\mu_u}{T}}+1},\qquad f_{\overline{u}}=\frac{1}{e^{\frac{\epsilon_u+\mu_u}{T}}+1}.
$$
Note that the difference between $\Phi_{ku}$ and $U_{ku}$ is due to the sign of the chemical potential $\mu_u$.  Analogous expressions are true for $d$ and $s$ quarks. 

When the density is high enough, $\mu>>T$ the emissivity described above may not the leading term anymore.
A possible energy loss process is due to the beta quark decay $u+e^{-}\to d+\nu_e$ or $d\to e^{-}+ u+\overline{\nu}_e$.
The neutrino emissivity in this case is given by the well known Iwamoto formula \cite{emissivity1}
$$
Q_\nu=\frac{914}{315}G^2_f \cos^2\theta_c\mu_\mu \mu_d \mu_e\alpha_s T^6.
$$
Here the Cabbibo angle $\theta_c$ is such that $\cos ^2\theta_c\sim 0.948$ and the condition of $\beta$ equilibrium is 
$$
\mu_u=\mu_d+\mu_e,\qquad \mu_s=\mu_d+\mu_e.
$$
For high densities, the following approximation is valid
$$
\mu_\mu=\mu_d, \qquad \mu_e=3^{\frac{1}{3}}Y_e^{\frac{1}{3}} \mu_u,
$$
where the number $Y_e$ for dense matter varies from $Y_e\sim 10^{-2}$ to $Y_e\sim 10^{-1}$.
In these terms the emissivity is given by
\be\lb{iwamoto}
Q_\nu=\frac{914}{315}G_f^2\mu^3 Y_e^{\frac{1}{3}}\alpha_s T^6.
\ee
There are other neutrino processes for matter at high densities that can be effective for cooling, examples can be seen in the references \cite{emissivity1}-\cite{emissivity12} and \cite{blaschke1}-\cite{blaschkesuppression}. It is important to remark that the formula (\ref{iwamoto})
does not assume that the CS phase takes place. In fact, there are phases of matter which are not represented as a quark-gluon plasma for which the emissivity may be strongly suppressed, examples are given in \cite{blaschkesuppression}.
In the following, it will be assumed that in the CS phase, the emissivity is very small in comparison with the afore mentioned processes. 

\subsection{An estimation of the emissivities}
The study of the emissivity $Q_\nu$ given in (\ref{pairemisivity}) requires an estimation of the integrals (\ref{termic}).
These integrals are all of the form
$$
I_{k}=\frac{1}{m^{k+3}\pi^2}\int_0^\infty \frac{\epsilon^k p^2 dp}{1+\exp(\frac{\epsilon-\mu}{T})},
$$
where the chemical potential $\mu$ can take positive and negative values and $k=2,1,0,-1$. As $\epsilon=\sqrt{p^2+m^2}$ for relativistic particles, it follows
that
$$
p^2dp=\epsilon\sqrt{\epsilon^2-m^2} \;d\epsilon\sim \epsilon^2\bigg(1-\frac{m^2}{2 \epsilon^2}-\frac{m^4}{8\epsilon^4}\bigg)d\epsilon.
$$
The integrals under consideration are then given by
$$
I_{k}=\frac{1}{ m^{k+3}\pi^2}\int_m^\infty \frac{\epsilon^{k+1}\sqrt{\epsilon^2-m^2}d\epsilon}{1+\exp(\frac{\epsilon-\mu}{T})}\simeq \frac{1}{m^{k+3}\pi^2}\int_m^\infty \frac{\epsilon^{k+2} d\epsilon}{1+\exp(\frac{\epsilon-\mu}{T})}\bigg(1-\frac{m^2}{2 \epsilon^2}-\frac{m^4}{8\epsilon^4}\bigg).
$$
Consider first the case $k=2$. The corresponding integral 
$$
I_{2}\simeq \frac{1}{m^{5}\pi^2}\int_{m}^\infty \frac{\epsilon^{4} d\epsilon}{1+\exp(\frac{\epsilon-\mu}{T})}\bigg(1-\frac{m^2}{2 \epsilon^2}-\frac{m^4}{8\epsilon^4}\bigg),
$$
can be found explicitly, the result is
$$
I_{2}\sim \frac{-1}{ m^{5}\pi^2}\bigg[24T^5\text{Li}_5(-e^{\frac{\mu-m}{T}})+24T^4 m \text{Li}_4(-e^{\frac{\mu-m}{T}})+12T^3m^2 \text{Li}_3(-e^{\frac{\mu-m}{T}})+4T^2 m^3 \text{Li}_2(-e^{\frac{\mu-m}{T}})
$$
$$
-T m^4\log(e^{\frac{\mu-m}{T}}+1)\bigg]+\frac{1}{2 m^{3}\pi^2}\bigg[2T^3 \text{Li}_3(-e^{\frac{\mu-m}{T}})+2T^2 m \text{Li}_2(-e^{\frac{\mu-m}{T}})
-T m^2\log(e^{\frac{\mu-m}{T}}+1)
\bigg]
$$
\be\lb{idos}
+\frac{1}{8\pi^2 m}[m-T \log(e^{\frac{\mu}{T}}+e^{\frac{m}{T}})].
\ee
Here the polylogarithm functions
\be\lb{poli}
\text{Li}_{1+s}(x)=\frac{1}{\Gamma(s+1)}\int_{0}^{\infty}\frac{x k^s dk}{e^{k}-x},
\ee
have been introduced. 

Unlike the previous case, the integrals corresponding to $k=1,0,-1$ are not explicit. For dealing with them, the following approximated scheme will be employed.
Consider the case $k=1$ namely
$$
I_{1}\simeq \frac{1}{m^{4}\pi^2}\int_{m}^\infty \frac{\epsilon^{3} d\epsilon}{1+\exp(\frac{\epsilon-\mu}{T})}\bigg(1-\frac{m^2}{2 \epsilon^2}-\frac{m^4}{8\epsilon^4}\bigg).
$$
The first two terms can be integrated explicitly but, to the best of our knowledge, there is no primitive for the last one. However, if one separates the last term and write
the last expression as
$$
I_{1}\simeq \frac{1}{m^{4}\pi^2}\int_{m}^\infty \frac{\epsilon^{3} d\epsilon}{1+\exp(\frac{\epsilon-\mu}{T})}\bigg(1-\frac{m^2}{2 \epsilon^2}\bigg)-\frac{1}{\pi^2}\int_{m}^\infty \frac{ d\epsilon}{1+\exp(\frac{\epsilon-\mu}{T})}\frac{1}{8\epsilon},
$$
then, by further making the variable change $\epsilon=T\log \eta$ one obtains
$$
I_{1}\simeq \frac{1}{m^{4}\pi^2}\int_{m}^\infty \frac{\epsilon^{3} d\epsilon}{1+\exp(\frac{\epsilon-\mu}{T})}\bigg(1-\frac{m^2}{2 \epsilon^2}\bigg)-\frac{e^{\frac{\mu}{T}}}{8\pi^2}\int_{e^{\frac{m}{T}}}^\infty \frac{ d\eta}{\eta(e^{\frac{\mu}{T}}+\eta)}\frac{1}{\log \eta}.
$$
The last integral can be successively be approximated by separating the cases $\mu<m$ or $\mu>m$. For instance, if $\mu<m$ the integrand can be expanded in terms of $e^{\frac{\mu}{T}}/\eta$ by use of geometric series, the result is
$$
I_{1}\simeq \frac{1}{m^{4}\pi^2}\int_{m}^\infty \frac{\epsilon^{3} d\epsilon}{1+\exp(\frac{\epsilon-\mu}{T})}\bigg(1-\frac{m^2}{2 \epsilon^2}\bigg)-\frac{e^{\frac{\mu}{T}}}{8\pi^2}\int_{e^{\frac{m}{T}}}^\infty \frac{d\eta}{\eta^2\log \eta}\bigg[1-\frac{e^{\frac{\mu}{T}}}{\eta}+\frac{e^{\frac{2\mu}{T}}}{\eta^2}\bigg]\qquad \text{for}\qquad \mu<m.
$$
In the other possible situation namely $\mu>m$ there are two $\eta$ regions to consider, the first corresponds to $\epsilon<\mu$ and the second to $\epsilon>\mu$.
In the first region the expansion parameter is  $\eta/e^{\frac{\mu}{T}}$ and in the second $e^{\frac{\mu}{T}}/\eta$. The resulting integral is
$$
  I_{1}\simeq \frac{1}{m^{4}\pi^2}\int_{m}^\infty \frac{\epsilon^{3} d\epsilon}{1+\exp(\frac{\epsilon-\mu}{T})}\bigg(1-\frac{m^2}{2 \epsilon^2}\bigg)-\frac{e^{\frac{\mu}{T}}}{8\pi^2}\int_{e^{\frac{\mu}{T}}}^\infty \frac{d\eta}{\eta^2\log \eta}\bigg[1-\frac{e^{\frac{\mu}{T}}}{\eta}+\frac{e^{\frac{2\mu}{T}}}{\eta^2}\bigg]  
$$
$$
  -\frac{1}{8\pi^2}\int_{e^{\frac{m}{T}}}^{e^{\frac{\mu}{T}}}\frac{d\eta}{\eta\log \eta}\bigg[1-\frac{\eta}{e^{\frac{\mu}{T}}}+\frac{\eta^2}{e^{\frac{2\mu}{T}}}\bigg],\qquad \text{for},\qquad \mu>m.
$$
The resulting integrals are explicit,  the result is
$$
I_{1}\sim \frac{-1}{ m^{4}\pi^2}\bigg[6T^4 \text{Li}_4(-e^{\frac{\mu-m}{T}})+6T^3m \text{Li}_3(-e^{\frac{\mu-m}{T}})+3T^2 m^2 \text{Li}_2(-e^{\frac{\mu-m}{T}})
-T m^3\log(e^{\frac{\mu-m}{T}}+1)\bigg]
$$
\be\lb{iuno1}
+\frac{1}{2 m^{2}\pi^2}\bigg[T^2 \text{Li}_2(-e^{\frac{\mu-m}{T}})-T m \log(e^{\frac{\mu-m}{T}}+1)
\bigg]+\frac{1}{8\pi^2}\bigg[e^{\frac{\mu}{T}} \text{Ei}\bigg(-\frac{m}{T}\bigg)-e^{\frac{2\mu}{T}} \text{Ei}\bigg(-\frac{2m}{T}\bigg)+e^{\frac{3\mu}{T}}\text{E}i\left(
-\frac{3m}{T}\right)\bigg],
\ee
for $\mu<m$ and
$$
I_{1}\sim \frac{-1}{ m^{4}\pi^2}\bigg[6T^4 \text{Li}_4(-e^{\frac{\mu-m}{T}})+6T^3m \text{Li}_3(-e^{\frac{\mu-m}{T}})+3T^2 m^2 \text{Li}_2(-e^{\frac{\mu-m}{T}})
-T m^3\log(e^{\frac{\mu-m}{T}}+1)\bigg]
$$
$$
+\frac{1}{2 m^{2}\pi^2}\bigg[T^2 \text{Li}_2(-e^{\frac{\mu-m}{T}})-T m \log(e^{\frac{\mu-m}{T}}+1)
\bigg]+\frac{1}{8\pi^2}\bigg[e^{\frac{\mu}{T}} \text{Ei}\bigg(-\frac{\mu}{T}\bigg)-e^{\frac{2\mu}{T}} \text{Ei}\bigg(-\frac{2\mu}{T}\bigg)+e^{\frac{3\mu}{T}}\text{E}i\left(
-\frac{3\mu}{T}\right)\bigg]
$$
\be\lb{iuno2}
+\frac{1}{8\pi^2}\bigg[e^{-\frac{2\mu}{T}}\text{Ei}\bigg(\frac{2m}{T}\bigg)-e^{-\frac{\mu}{T}}\text{li}(e^{\frac{m}{T}})+\log\bigg(\frac{m}{\mu}\bigg)-e^{-\frac{2\mu}{T}}\text{Ei}\bigg(\frac{2\mu}{T}\bigg)+e^{-\frac{\mu}{T}}\text{li}(e^{\frac{\mu}{T}})\bigg],
\ee
for $\mu>m$. In the above expressions the exponential integral function
\be\lb{ei}
\text{Ei}(x)=-\int_{-x}^\infty \frac{e^{-t}dt}{t},
\ee
and the logarithmic integral
\be\lb{li}
\text{li}(x)=-\int_{0}^{x}\frac{dt}{\log t},\qquad \text{for $x<$1},\qquad \text{li}(x)=-\text{PV}\int_{0}^{x}\frac{dt}{\log t},\qquad \text{for $x>1$.}
\ee
were introduced. Here PV denotes the Cauchy principal value, and  $x=1$ is a singular value. 

The remaining integrals $k=0,-1$ can be approximated by exactly the same method. Without quoting the details, the result is
$$
I_{0}\sim \frac{-1}{m^{3}\pi^2}\bigg[2T^3 \text{Li}_3(-e^{\frac{\mu-m}{T}})+2T^2 m \text{Li}_2(-e^{\frac{\mu-m}{T}})
-T m^2\log(e^{\frac{\mu-m}{T}}+1)
\bigg]
+\frac{1}{2 m\pi^2}[m-T\log(e^{\frac{\mu}{T}}+e^{\frac{m}{T}})]
$$
\be\lb{icero1}
-\frac{m}{8\pi^2T}\bigg[e^{\frac{\mu}{T}} \text{Ei}\bigg(-\frac{m}{T}\bigg)-2e^{\frac{2\mu}{T}} \text{Ei}\bigg(-\frac{2m}{T}\bigg)+3e^{\frac{3\mu}{T}}\text{E}i\bigg(
-\frac{3m}{T}\bigg)\bigg]
-\frac{1}{8\pi^2}\bigg[e^{\frac{\mu-m}{T}}-e^{\frac{2\mu-2m}{T}}+e^{\frac{3\mu-3m}{T}}\bigg],
\ee
$$
I_{-1}\sim \frac{-1}{m^{2}\pi^2}[T^2 \text{Li}_2(-e^{\frac{\mu-m}{T}})-T m \log(e^{\frac{\mu-m}{T}}+1)]+\frac{1}{2\pi^2}\bigg[e^{\frac{\mu}{T}} \text{Ei}\bigg(-\frac{m}{T}\bigg)-e^{\frac{2\mu}{T}} \text{Ei}\bigg(-\frac{2m}{T}\bigg)+e^{\frac{3\mu}{T}}\text{E}i\left(
-\frac{3m}{T}\right)\bigg]
$$
$$
+\frac{m^2}{8T^2\pi^2}\bigg[\frac{e^{\frac{\mu}{T}}}{2} \text{Ei}\bigg(-\frac{m}{T}\bigg)-2e^{\frac{2\mu}{T}} \text{Ei}\bigg(-\frac{2m}{T}\bigg)+\frac{9}{2}e^{\frac{3\mu}{T}}\text{E}i\bigg(
-\frac{3m}{T}\bigg)\bigg]
+\frac{m}{8T\pi^2}\bigg[\frac{1}{2}e^{\frac{\mu-m}{T}}-e^{\frac{2\mu-2m}{T}}+\frac{3}{2}e^{\frac{3\mu-3m}{T}}\bigg]
$$
\be\lb{imenos1}
+\frac{1}{16\pi^2}\bigg[-e^{\frac{\mu-m}{T}}+e^{\frac{2\mu-2m}{T}}-e^{\frac{3\mu-3m}{T}}\bigg],
\ee
for $\mu<m$ and
$$
I_{0}\sim \frac{-1}{m^{3}\pi^2}\bigg[2T^3 \text{Li}_3(-e^{\frac{\mu-m}{T}})+2T^2 m \text{Li}_2(-e^{\frac{\mu-m}{T}})
-T m^2\log(e^{\frac{\mu-m}{T}}+1)
\bigg]
+\frac{1}{2 m\pi^2}[m-T\log(1+e^{\frac{m-\mu}{T}})]
$$
$$
-\frac{m}{8\pi^2T}\bigg[e^{\frac{\mu}{T}} \text{Ei}\bigg(-\frac{\mu}{T}\bigg)-2e^{\frac{2\mu}{T}} \text{Ei}\bigg(-\frac{2\mu}{T}\bigg)+3e^{\frac{3\mu}{T}}\text{E}i\bigg(
-\frac{3\mu}{T}\bigg)\bigg]
+\frac{m}{8\pi^2 T}\bigg[2 e^{-\frac{2\mu}{T}} \text{Ei}\bigg(\frac{2m}{T}\bigg)
$$
\be\lb{icero2}
-2 e^{-\frac{2\mu}{T}} \text{Ei}\bigg(\frac{2\mu}{T}\bigg)+e^{-\frac{\mu}{T}}\text{li}(e^{\frac{\mu}{T}})-e^{-\frac{\mu}{T}}\text{li}(e^{\frac{m}{T}})\bigg]+\frac{1}{8\pi^2}[1-e^{\frac{m-\mu}{T}}+e^{\frac{2m-2\mu}{T}}],
\ee
$$
I_{-1}\sim \frac{-1}{m^{2}\pi^2}[T^2 \text{Li}_2(-e^{\frac{\mu-m}{T}})-T m \log(e^{\frac{\mu-m}{T}}+1)]
+\frac{1}{2\pi^2}\bigg[e^{\frac{\mu}{T}} \text{Ei}\bigg(-\frac{\mu}{T}\bigg)-e^{\frac{2\mu}{T}} \text{Ei}\bigg(-\frac{2\mu}{T}\bigg)+e^{\frac{3\mu}{T}}\text{E}i\left(
-\frac{3\mu}{T}\right)\bigg]
$$
$$
+\frac{m^2}{8T^2\pi^2}\bigg[\frac{e^{\frac{\mu}{T}}}{2} \text{Ei}\bigg(-\frac{\mu}{T}\bigg)-2e^{\frac{2\mu}{T}} \text{Ei}\bigg(-\frac{2\mu}{T}\bigg)+\frac{9}{2}e^{\frac{3\mu}{T}}\text{E}i\bigg(
-\frac{3\mu}{T}\bigg)\bigg]
+\frac{1}{2\pi^2}\bigg[e^{-\frac{2\mu}{T}}\text{Ei}\bigg(\frac{2m}{T}\bigg)-e^{-\frac{\mu}{T}}\text{li}(e^{\frac{m}{T}})
$$
$$
+\log\bigg(\frac{m}{\mu}\bigg)
-e^{-\frac{2\mu}{T}}\text{Ei}\bigg(\frac{2\mu}{T}\bigg)+e^{-\frac{\mu}{T}}\text{li}(e^{\frac{\mu}{T}})\bigg]
+\frac{m^2}{8\pi^2 T^2}\bigg[-2 e^{-\frac{2\mu}{T}}\text{Ei}\bigg(\frac{2\mu}{T}\bigg)+2 e^{-\frac{2\mu}{T}}\text{Ei}\bigg(\frac{2m}{T}\bigg)+\frac{e^{-\frac{\mu}{T}}}{2}\text{li}(e^{\frac{\mu}{T}})
$$
\be\lb{imenos2}
-\frac{e^{-\frac{\mu}{T}}}{2}\text{li}(e^{\frac{m}{T}})\bigg]+\frac{3m^2}{16\pi^2T \mu \pi^2}-\frac{1}{16\pi^2}(1-e^{\frac{m-\mu}{T}}+e^{\frac{2m-2\mu}{T}})+\frac{m}{16\pi^2 T}(e^{\frac{m-\mu}{T}}-2e^{\frac{2m-2\mu}{T}}).
\ee
for $\mu>m$. In these terms the emissivities (\ref{pairemisivity}) can be calculated by identifying $\Phi_k$ with $I_k$ with $\mu<0<m$ and $U_k$ with one of the $I_k$ depending
if $\mu>m$ or $\mu<m$.

\begin{figure}[h!]
\centering
\includegraphics[width=0.5\textwidth]{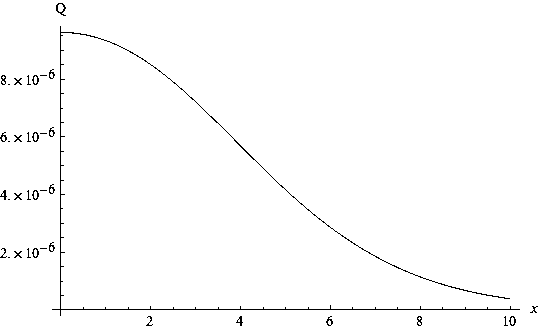}
$$
$$
\caption{The massless limit of $Q$ as a function of $x=\mu/T$ for $T=100$ MeV. The units in all the figures are MeV$^5$. }
\label{figura1}
\end{figure}
 
The formulas (\ref{idos})-(\ref{imenos2}) are approximated to a given order. However the massless limit $m\to 0$ is exact. In the massless regime the emissivities (\ref{pairemisivity}) are given by
\be\lb{emiteu}
Q\sim\frac{32G^2_f T^9}{\pi^5}A_{u+}^2[\text{Li}_5(-e^{\frac{\mu}{T}})\text{Li}_4(-e^{-\frac{\mu}{T}})+\text{Li}_5(-e^{-\frac{\mu}{T}})\text{Li}_4(-e^{\frac{\mu}{T}})].
\ee
As the figure 1 shows, this function is bell shaped and has values that run from $10^{-5}$ MeV$^5$ to $8 .10^{-6}$ MeV$^5$ for $T\sim 100$ MeV and $0<\mu<10$ T.  The units of the figures are all MeV$^5$. addition, for $\mu\to \infty$ and $T$ fixed it goes to zero. This makes sense, as the excess of particles over anti-particles
$$
n_q-n_{\overline{q}}\sim \frac{\mu}{T}\bigg[1+\frac{\mu^2}{\pi^2T^2}\bigg],
$$
is very large in this limit, so annihilation is likely be suppressed.
\begin{figure}[h!]
\centering
\includegraphics[width=0.5\textwidth]{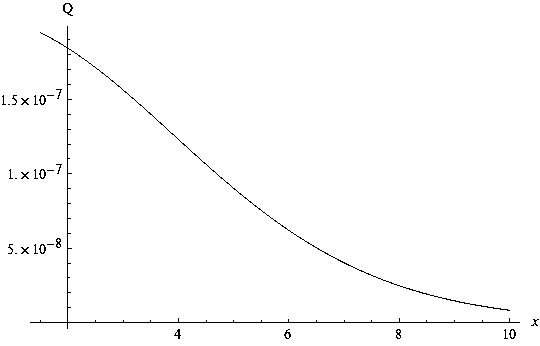}
$$
$$
\caption{The values of $Q$ for $m\sim m_u$ and $T=100$ MeV as a function of $x=\mu/T$.}
\label{figura2}
\end{figure}
It is interesting to analyse how the emissivity varies as a function of the quark mass $m$. It is not clear at first sight if the emissivity would grow or decay when $m$ increases, since the density of heavier particles is suppressed by a Fermi-Dirac factor but its decay rate seems to increase with the mass. 
\begin{figure}[h!]
\centering
\includegraphics[width=0.5\textwidth]{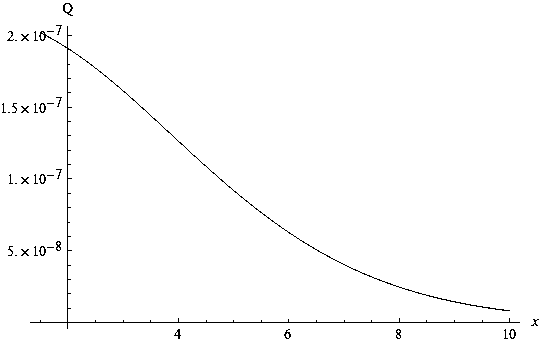}
$$
$$
\caption{The values of $Q$ for $m\sim m_s$ and $T=100$ MeV as a function of $x=\mu/T$.}
\label{figura3}
\end{figure}
 At the end, it is expected that $Q\to 0$ when $m\to \infty$, as such a massive quark is likely to decay fast but it strongly suppressed by thermodynamics. The exponential integral $\text{Ei}(-x)$ is such that $\text{Ei}(-x)\to \infty$ when $x\to 0$.  However, $x^\alpha \text{Ei}(-x)\to 0$
in this limit. In addition $x^\alpha \text{Ei}(-x)\to 0$ when $x\to \infty$ and $\alpha\geq 1$. The same result is true for the logarithm or polylogarithm terms appearing in (\ref{iuno1})-(\ref{imenos2}). By use of these facts, one may obtain that
$$
\lim_{m\to \infty} Q(m, \mu, T)\to 0.
$$
This is the expected result. On the other hand, there is an essential singularity when $\mu\to\infty$ and $m\to \infty$, since the behaviour of $Q$ depends on the curve $(m(t), \mu(t))$ chosen for taking the limit.  
\begin{figure}[h!]
\centering
\includegraphics[width=0.5\textwidth]{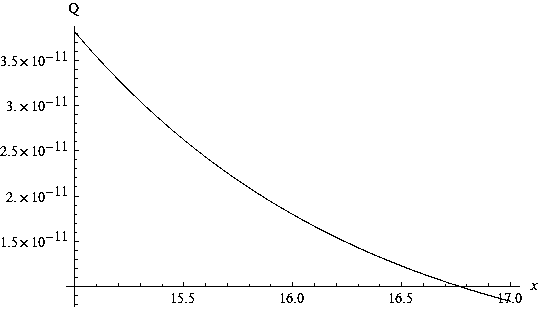}
$$
$$
\caption{The values of $Q$ for a very massive quark with mass $m\sim 10 m_s$ and $T=100$ MeV as a function of $x=\mu/T$ in the range $15<x<20$. Its values are never negative.}
\label{figura4}
\end{figure}

The conclusion given above related to the asymptotic behaviour of $Q$ with respect to the mass however,  does not give relevant information for moderate values of $m$ such as $m\sim T$ or $m\sim m_u\sim m_d$. These moderate values are the ones
relevant for the present work. In order to understand the behaviour of emissivity for these mass values, a numerical estimation is in order. We have plotted with Mathematica the emissivity in several regimes. We collect here some relevant cases for illustrative purposes. Figure 2 shows that for a light quark with mass $m_u$ the emissivity is not considerably deviated from the massless case of figure 1. Figure 3 shows that for a quark of mass $m_s\sim T$
the deviation is also not very significant. We have plotted the emissivities for other temperatures and we have found a similar behaviour. A significant variation appears when the quark mass $m$ is considerably larger than the temperature $T$, as figure 4 shows. In this case the emissivity gets significantly suppressed. Our results suggest that the emissivities for particles with masses below or of the order of the temperature $T$ are more or less similar, but when the mass $m>>T$ then $Q$ starts to decrease considerably. The emissivities plotted above are all related to the case $\mu>m$, but in the other regime a similar conclusion applies. For this reason, for the present problem in consideration, the massless emissivity will be considered.
In other words, the emissivity will be given by
\be\lb{emo}
Q\sim\frac{96G^2_f T^9}{\pi^5} F\bigg(\frac{\mu}{T}\bigg),
\ee
with $F(x)$ a function taking values between $1$ and $0.1$ when $0<x<10$.

\section{Description of the evolution of the bubble}

\subsection{The evolution of an small nugget}
After the emissivity has been characterised, the next section is to describe qualitatively the evolution of the bubble by use of the baryon number conservation condition (\ref{ecomocos3}) together with
the equations of motion (\ref{segun})-(\ref{prime}).
As discussed in section 2, the initial radius is assumed to be of order $R_0\sim 1$ cm. In addition, the small choice $N\sim 1-10$ in (\ref{ecomocos3}) will be employed. 
 This corresponds to a baryon number $B\sim 10^{28}-10^{29}$ for the object. For radius $R<1$ cm and temperatures $T<100$ MeV the terms proportional to $\sigma R$ predominate over the ones proportional
 to $R^2 T^4$, and the emissivity $Q_\nu$ is even smaller. The difference becomes more accentuated as the radius or the temperature decreases. This means that the surface tension $\sigma$ plays a major role in contracting the object initially.
 This situation, as discussed below, is reversed when the chemical potential reaches values $\mu>>T$.

 If initially the chemical potential $\mu=0$, then the term proportional to $R^3$ in (\ref{ecomocos3}) vanishes. However, by taking into account that $T_0\sim 100$ MeV it is seen
 that $R_0 T_0\sim 10^{13}$. This suggests that the volume term in  (\ref{ecomocos3}), which is proportional to $R^3T^2\mu$ may quickly starts to predominate over the surface term proportional to $R^2T^2$, even when $\mu<<T$.
 This follows from the fact that $N\sim 1-10$ is considerably smaller than $R_0 T_0\sim 10^{13}$. 
 Now, in the regime $\mu<<T$ it may be assumed in  (\ref{segun}) that
\be\lb{chuco}
1+\frac{30\mu^2}{7\pi^2T^2}+\frac{15\mu^4}{7\pi^4T^4}\sim 1.
\ee
This does not mean that the time derivatives of these quantities are small, since even a small function may have a large slope at some point. But
if it is supposed that the slope is moderate then equation (\ref{segun}) implies that
$$
\frac{28\pi^3 g T^3}{270}\dot{T}+\frac{4\pi}{3} Q_\nu\geq 0.
$$
This follows from the fact that the first term in (\ref{segun}) is assumed to be negative. This is justified because the surface tension term proportional $\sigma$ is large in comparison with the other components of the pressure, for a radius smaller than a centimeter. Thus it may be assumed safely that the bubble is initially contracting and perhaps cooling, which means that the derivatives of all these terms are all negative. By use of (\ref{emiteu}) and the numerical values of the emissivity found
in the previous section, the last equation integrates approximately to
$$
T\geq \frac{T_0}{\bigg[1+10^{10}\bigg(\frac{t}{\text{s}}\bigg)\bigg]^{\frac{1}{5}}}.
$$
Here the functional form (\ref{emo}) the value $G_f\sim 10^{-5}$GeV$^{-2}$ were taken into account, and the chemical potential $\mu$ in (\ref{emiteu}) was set to zero for simplicity. 

The last formula shows that, for the temperature to lower down to $T=T_0/3$, at least a  time $t\sim10^{-9}$s is required, irrespective to the initial temperature value $T_0$. This is exactly the time if no contraction takes place, that is, for constant $R$.
For decreasing $R$ this value may be larger. However, we will assume that this is the characteristic time such that,
for $t\leq 10^{-9}$s, the temperature remains constant and after that, a considerable cooling starts due to the neutrino emission. Note that the external temperature (\ref{expres}) will be also constant for such short time period. 
In the present scheme, there is no identification between the temperature $T$ of the object and the external temperature $T_e$.  This line of reasoning may not be true when $\mu>>T$, as (\ref{chuco}) would not be valid, and the presence of large derivatives
of $\mu$ may slow down the cooling. However, it is reasonable to assume that for a time of $t\sim 10^{-4}$s, which is five orders of magnitude larger,  the temperature will be of the order $T_0/3$ or smaller. This, as will be discussed below, will imply that the object falls in the CFL phase \cite{cfl}-\cite{cfl2}. 

All the previous approximation is assumed to be valid for small chemical potential $\mu<<T$. By taking into account 
(\ref{ecomocos3}) and by neglecting the surface term, it follows that
$$
 R^3\mu^2T=3\pi^4 R_0^2T_0^2.
$$
This suggest that the regime $\mu\sim T$ is achieved for some radius $R_1$ and some temperature and chemical potential $\mu_1\simeq T_1$ such that 
$$
R_1 T_1\sim (R_0T_0)^{\frac{2}{3}}.
$$
Since initially $R_0 T_0\sim 10^{13}$, it is seen that now $R_1\mu_1\sim 10^{10}$. If the temperature is not significantly changed, then this implies that there is a violent contraction to $R_1\sim 10^{-2}-10^{-3}$cm
and the chemical potential reaches the value $\mu \sim 100$ MeV.  In order to check if this is true note that, at initial stages, the dominant term in the right hand side of (\ref{prime}) is the one proportional to the surface tension $\sigma$. In this equation, this terms dominates the emissivity, whose maximum value corresponds to $T\sim 100$ MeV and $R\sim1$ cm.  If all the non relevant terms are neglected in (\ref{prime}) the equation simplifies to a 
Newtons law equation of the form
$$
R^2 \ddot{R}\sim -2R.
$$
The integration of this equation gives that
$$
\frac{\sqrt{\pi}R_0}{2} \text{Erf}\bigg(\sqrt{\log\frac{R_0}{R}}\bigg)\sim t.
$$
Here $\text{Erf}(x)$ is the error function. For a contraction of $R=10^{-3}R_0$, this gives around $t\sim 10^{-10}$s.  However, these arguments have a problem, as the velocity of the wall is given by
$$
\dot{R}^2=-4 \log \frac{R}{R_0}.
$$
If this expression is taken literally into account, then  contraction velocity reaches a superluminal value $\dot{R}^2>1$ at some point. This suggest the the bubble surface may reach velocities close to light, and the non relativistic approximation employed
here is not valid. However, assume that the bubble wall moves with light velocity. Then, it makes around $10^{9}$ m per second, which means that it travels a distance of the order of centimeter with a time around $t\sim 10^{-11}$s.
All the previous discussion suggests that a contraction from $R_0$ to $R=10^{-3}R_0$ occurs in a time of the order $t\sim 10^{-9}-10^{-11}$s. In this period
$\mu$ reaches the value $\mu\sim T_0$ and, as the process seems to be very quickly, it is plausible that no significant cooling takes place during this contraction. Thus, at the end $\mu\sim T_0\sim 100$ MeV.

Once the chemical potential reaches the value $\mu\sim 100$ MeV, there is  a further period of contraction in which the chemical potential grows. 
In order to see this, note that for $\mu>>T_0$ the conservation of baryon number 
(\ref{ecomocos3}) gives that
\be\lb{putin}
\mu\sim  \frac{2(R_0T_0)^{\frac{2}{3}}}{R}.
\ee
The last formula shows that
$\mu=\mu(R)$, and it is independent on the value of $T$. On the other hand, the right side of (\ref{prime}) can be approximated as
$$
\Delta P=-8\pi \sigma R +4\pi R^2\frac{7}{8}\frac{\pi^2 g T^4}{90}\bigg(1+\frac{30\mu^2}{7\pi^2T^2}+\frac{15\mu^4}{7\pi^4T^4}\bigg)-4\pi R^2E_B \Theta(\mu-\mu_1)\bigg(1-\frac{\mu_1^2}{\mu^2}\bigg)
$$
$$
-4\pi R^2\frac{7}{8}\frac{\pi^2 g T_e^4}{90}+F_\eta
\sim -8\pi \sigma R + \frac{2^4\;g}{12\pi}\frac{(R_0T_0)^{\frac{8}{3}}}{R^2}.
$$
In making this approximation, the term proportional to $\mu^4/T^4$ was assumed to be leading, and the formula (\ref{putin}) was employed for replacing $\mu$ as a function of $R$. This term clearly is larger than $\mu^2/T^2$
and than one. The terms proportional to $E_B\sim (150\;\text{MeV})^4$ and to $T_e$ are also small in comparison with $\mu>>100$ MeV, for a radius smaller than a centimeter. Now, if the object is assumed to enter in the CFL phase, the emissivity may be neglected.
The equilibrium position $R_e$ is then the zero of the last expression and, with the present choice of parameters, is given by
$$
R_e\sim 4.\;10^{-6} \text{cm}.
$$
A further comment about this magnitude is in order. The surface tension $\sigma$ of the domain wall, as discussed in (\ref{sigma}), is $\sigma\sim 10^{20}$ MeV$^3$. However, for small bubbles,
a radial dependence $\sigma=\sigma(R)$ may appear. In obtaining this number, we have assumed that $\sigma\sim 10^{19}$MeV$^3$, that is an order of magnitude less than the original value. 
This difference does not change significantly the equilibrium radius $R_e$, it simply corrects it by a factor of two.

The chemical potential $\mu$ that follows from (\ref{putin}) is indeed very large, $\mu>500$ MeV. It is also consistent with \cite{zhitnitsky12} when chemical potential indeed assumes the value well above $400$ MeV during a time scale of the order $10^{-4}$s.  Thus the hypothesis that the emissivity can be safely omitted is reasonable.
In these terms, one has from (\ref{masamasa}) that at the end of the evolution  $M\sim10^{32}$MeV. The baryon number is $B\sim 10^{29}$. This leads to an energy per baryon
$M/B\sim$ GeV, which is of the order of a typical nucleon $m_N$ formed during that epoch. This is a condition  for warrant the stability of the object \cite{zhitnitsky1}.

The object then makes oscillations around the equilibrium position. The linealization of (\ref{prime}) around this equilibrium position $R_e$ gives
$$
\delta\ddot{R}+\frac{2}{\tau}\delta\dot{ R}+\omega^2 \delta R=0.
$$
This equation corresponds to exponentially damped oscillator with characteristic time scale $\tau$ and frequency $\omega^2$ which, in this case, are given by
\be\lb{nomers}
\tau\sim \frac{10\pi\sigma R_e}{3\eta},\qquad \omega^2\sim \frac{18}{5 R_e^2}.
\ee
With the values employed in this section, it follows that $\tau \sim 10^{-4}-10^{-3}$s and  $\omega \tau\sim 10^{13}\sim m_\pi/m_a$. This implies that the external temperature $T_e$ in (\ref{expres})
is close to $T_e\sim T_0/\sqrt{10}\sim T_0/3$. Thus, it is plausible that the external temperature when the object is formed is close to the value $T\sim 41$ MeV, a number that have 
several interesting phenomenological consequences.  

There should be however some words about the linealization performed here.  As pointed out in previous paragraphs, there may be a dependence between the surface tension and the bubble radius, that is, $\sigma=\sigma(R)$.The value of $R_e$ obtained here is two or three orders of magnitude smaller than the one obtained in \cite{zhitnitsky6}, but the value of $\sigma$ employed here
is two of three orders of magnitude larger. This implies that the time of formation $\tau$ is approximately the same of that reference. It should be remarked however, that 
the value of $\eta$ was calculated by assuming baryon number equal to zero, thus a more precise knowledge of this coefficient of course is desirable.

\subsection{A comment about axion emission}
The evolution considered in the previous subsections did not took into account that a contracting axion wall should emit axions \cite{sikivie2}-\cite{sikivie3}. The axion emission may play a role analogous to neutrino emissivity. For a wall with tension $\sigma$ alone, the equations of motion resulting by considering its interaction with the axion primordial soup would be
$$
\frac{d\sigma R^2}{dt}=-\widetilde{\rho}_a(t) R^2,
$$
with
\be\lb{axionoss}
\widetilde{\rho}_a(t)=10^{-9} m_\pi^2 f_\pi^2 \bigg(\frac{10^{-4}\text{seg}}{t}\bigg)^{\frac{3}{2}}\frac{v}{10^{10} \text{GeV}}.
\ee
This equation would be right for an empty bubble, but in the present work the bubble is in addition emitting neutrinos and entering into the color-flavor locked phase.
It is important to compare the effect of a term $\rho_a(t) R^2$ with the neutrino emissivity. Here the density $\rho_a(t)$ is not identified with (\ref{axionoss}) and parameterize our ignorance about the details of the emission. By taking into account (\ref{emo}), it follows that for $R\sim 10^2$ m the inequality
$$
\rho_a R^2>\frac{4\pi R^3}{3} Q_\nu,
$$
is satisfied for $\rho_a\geq 10^{10} \widetilde{\rho}_a$, which is an enormous density. Only at $R\sim 10^{-6}$ cm and at density (\ref{axionoss}) the axion emission is comparable
with the emissivity but, at these stages, the pressure terms are already more important that both emission terms.  For this reason these terms were neglected in the dynamics, as it may complicate the analysis qualitatively, but not give rise to large deviation from the behavior just described.

\section{A further approximation}

In the previous section, it was assumed that the internal temperature of the quark nugget $T_i$ may differ from the external one $T_e$. But after employing the assumed approximations, it was found that a large part of the bubble evolution takes place at almost constant temperature. At these stages of the universe evolution however, the density of the surrounding electrons, positrons, baryons, photons is very high. It is likely that the corresponding time scale for the thermal equilibration is much shorter than the neutrino-induced cooling. Thus, the internal temperature of the object and universe temperature can be matched at these epochs. This is until the temperature drops below the Big Bang Nucleosynthesis era. At these times, if  the object falls into the Colour-Superconducting phase, it is expected the neutrino emissivity to drop. It is likely that only at those stages the object has its own temperature, independent on the environment one.

If the matching of temperatures is assumed to hold, then the equations (\ref{ecomoco2})-(\ref{ecomoco}) has to be modified. If one assumes that the internal temperature is equal to the external one, then one of these equations or a combination of them has to be deleted. The energy balance equation (\ref{ecomoco2}) is likely to include terms related to the baryons and the universe expansion, and will  give the final result that $T_i=T_e$. Thus, this equation will be ignored, as it is not properly including the effect of the external fluxes. The Newton type equation (\ref{ecomoco}) will be kept, and will be written in the following form
$$
\frac{d}{dt}\bigg\{\bigg[4\pi \sigma R^2+\frac{4\pi R^3}{3}\frac{7}{8}\frac{\pi^2 g T^4}{30}\bigg(1+\frac{30\mu^2}{7\pi^2T^2}+\frac{15\mu^4}{7\pi^4T^4}\bigg)+\frac{4\pi R^3}{3}E_B \Theta(\mu-\mu_1)\bigg(1-\frac{\mu_1^2}{\mu^2}\bigg)\bigg]\dot{R}\bigg\}+\frac{4\pi}{3} R^3 Q_\nu
$$
$$
=-8\pi \sigma R +4\pi R^2\frac{7}{8}\frac{\pi^2 g T^4}{90}\bigg(1+\frac{30\mu^2}{7\pi^2T^2}+\frac{15\mu^4}{7\pi^4T^4}\bigg)-4\pi R^2E_B \Theta(\mu-\mu_1)\bigg(1-\frac{\mu_1^2}{\mu^2}\bigg)-4\pi R^2\frac{7}{8}\frac{\pi^2 g T_e^4}{90}+F_\eta.
$$
By expanding the derivatives properly, it leads to the expression
$$
\bigg[4\pi \sigma+ \frac{2^4 g}{12\pi}\frac{(R_0T_0)^{\frac{8}{3}}}{R^3}+\frac{4\pi R}{3}E_B \Theta(\mu-\mu_1)\bigg(1-\frac{\mu_1^2 R^2}{4(R_0T_0)^{\frac{4}{3}}}\bigg)\bigg]\ddot{R}
+\bigg[8\pi \sigma - \frac{2^4 g}{12\pi}\frac{(R_0T_0)^{\frac{8}{3}}}{R^3}
$$
$$
+\bigg(4\pi RE_B \Theta(\mu-\mu_1)+\frac{8\pi R(R_0 T_0)^{\frac{2}{3}}E_B}{3} \delta(\mu-\mu_1)\bigg)\bigg(1-\frac{\mu_1^2 R^2}{4(R_0T_0)^{\frac{4}{3}}}\bigg)-\frac{4\pi R^3 \mu_1^2 E_B}{3(R_0 T_0)^{\frac{4}{3}}}\Theta(\mu-\mu_1)\bigg]\frac{\dot{R}^2}{R}
$$
$$
+\frac{4\pi}{3} R Q_\nu=-\frac{8\pi \sigma}{R} + \frac{2^4 g}{12\pi}\frac{(R_0T_0)^{\frac{8}{3}}}{R^4}-4\pi E_B \Theta(\mu-\mu_1)\bigg(1-\frac{\mu_1^2 R^2}{4(R_0T_0)^{\frac{4}{3}}}\bigg)+F_\eta.
$$
The linealization of this equation gives essentially the same numbers as (\ref{nomers}). Thus, the size of the object and the time of formation is not considerably changed by the matching of the internal and external temperature. 

A final comment about the linealization is in order. The full non linear equation described above, and the one in the previous section, contain scales with very different magnitude orders. For instance, by use the numbers obtained in the present work, the previous equation may be expressed in dimensionless form as 
\be\lb{adi}
m^2 \bigg[1+ \frac{1}{10^{26}x^3}\bigg]\ddot{x}
+m^2\bigg[2- \frac{1}{10^{26}x^3}\bigg]\frac{\dot{x}^2}{x}
+30 x =-\frac{2}{x}+ \frac{10^{-26}}{x^4}
-2.10^{-15} \frac{m\dot{x}}{x},
\ee
with $m$ denoting a meter and $x=R m^{-1}$. This shows the different scales that are involved in the problem. 
The behavior of this  type of equations were considered in \cite{zhitnitsky} numerically. The type of equations considered in that reference are of the form
$$
\sigma(R) \ddot{R}=-\frac{2\sigma(R)}{R}-\frac{\sigma \dot{R}^2}{R}-4\eta\frac{\dot{R}}{R}+\alpha T^4\bigg[\beta+\gamma\bigg(\frac{R_0}{R}\bigg)^2+\frac{1}{72}\bigg(\frac{R_0}{R}\bigg)^4\bigg].
$$
Here $\alpha$, $\beta$ and $\gamma$ are constants, which may have very different  orders of magnitude.
The equation (\ref{adi}) is of this type, as the right hand side contain terms that goes as $R^{-2}$ and $R^{-4}$ and to $\dot{R}/R$. It is difficult to see the oscillations numerically due to the discrepant scales involved, but they have been seen in \cite{zhitnitsky6} by properly scaling some parameters of the models. These numerical studies support the linealization procedure described in (\ref{nomers}), at least at qualitative level.

 \section{Viability of axion nuggets as dark matter candidates}

In the present work, the formation of axion quark nuggets and their possible evolution was analysed, but taking into account a large emission of neutrinos inside the object.
As a result, the final radius of the object is smaller than the one predicted in \cite{zhitnitsky4}. However, the analysis done here employs a smaller axion mass and the time of formation
of the object remains basically the same as in \cite{zhitnitsky}. 
The universe external temperature at the formation time is around $T_f\sim 41$ MeV.  The importance of this value is the following \cite{zhitnitsky1}. The baryon to photon density that is currently available is $$\eta\sim\frac{n_B-\overline{n}_B}{n_\gamma}\sim 6\cdot 10^{-10}.$$ This value can estimated as $\eta\sim T_{eq}/m_N$ between the temperature $T_{eq}\sim1$ eV of equilibrium between radiation and matter. Here $m_N$ is a typical nucleon mass,
which, in order to reproduce the observations, should be $m_N\sim$1 GeV. On the other hand, at the time of formation $n_B-\overline{n}_B\sim n_{B}\sim e^{-\frac{m_N}{T_f}}$.
The value $T_f\sim 41$ MeV is the one consistent with the measurements of the parameter $\eta\sim 6\;10^{-10}$.  
For this reason, the possibility that these objects are a considerable fraction of dark matter and may contain a large number of anti-baryon number, consistent
with the measured baryon-anti baryon asymmetry, remains plausible.

There are several consistency checks that these objects should fulfill in order to classify as cold dark matter candidates \cite{zhitnitsky}-\cite{zhitnitsky13}. It is important to see
if these conditions are satisfied by the objects described in the present work. 
\\

\emph{The axion quark nuggets are long lived}
\\

The first condition is that these objects  should be long lived.
 By assuming a geometric cross section, the total number of collisions between ordinary hadrons and axion quark nuggets is given by
$$
\frac{dW}{dt}=4\pi R^2 n_B v,\qquad n_B\sim\frac{0.15\rho_{dm}}{\text{GeV}}.
$$
From here it follows that the annihilation of baryon charge until the present time is
$$
\Delta B=\frac{1}{H}\frac{dW}{dt}.
$$
By employing a age of the universe of the order $t_u\sim 10^{17}$s, a typical hadron velocity $v\sim 10^{-3}$ and the formation size of the object $R_e\sim 10^{-5}$cm, it follows that $\Delta B\sim 10^{17}$, which is a value less
than the value $B\sim 10^{28}-10^{29}$ employed here. For larger values of $R_e$ and $B$, the same conclusion holds. Thus, these objects have a life time larger
than the universe age. 
\\

\emph{The energy density contribution of the nuggets}
\\

Another important aspect is the ratio between the energy density contribution of quark nuggets and ordinary baryons \cite{zhitnitsky2}. If one assumes
that these nuggets are the most important component of dark matter, then the excess of anti-baryons is hidden inside these compact objects and is of the order
of the baryon excess
$$
\overline{n}_{\text{dm}}-n_{\text{dm}}=\frac{1}{B}(n_B-\overline{n}_B)\sim\frac{n_B}{B}.
$$
By assuming that the excess of anti-nuggets is of the same order than the number of nuggets and anti-nuggets, one has then 
that $n_{\text{dm}}+\overline{n}_{\text{dm}}=C(\overline{n}_{\text{dm}}-n_{\text{dm}})$ with $C\geq 1$. This, combined with the previous
relation, gives that
$$
\frac{\overline{n}_{\text{dm}}+n_{\text{dm}}}{n_B}=\frac{C(\overline{n}_{\text{dm}}-n_{\text{dm}})}{n_B}= \frac{C}{B}.
$$
In these terms it follows that
\be\lb{magic}
\frac{\Omega_{\text{dm}}}{\Omega_B}\sim\frac{m_{\text{qn}}(\overline{n}_{\text{dm}}+n_{\text{dm}})}{m_N n_B}\sim\frac{C m_{\text{qn}}}{B m_N}.
\ee
By taking into account that $m_{\text{qn}}\sim B m_N$, one has that $\Omega_{\text{dm}}\geq \Omega_{B}$, within a magnitude order. This is an interesting feature,
since this relation is difficult to establish for models of dark matter not related to ordinary quark or baryon degrees of freedom \cite{zhitnitsky2}.
\\

\emph{The interaction with photons}
\\

It should be emphasized that the axion quark nuggets do interact with photons \cite{zhitnitsky5}. However, this does not pose 
a problem for being cold dark matter candidates if the mean free time for a photon to encounter a nugget is larger than the age of the universe. The mean free time for photons to collide
with a nugget is given by
$t_h=(n_B\sigma)^{-1}$,
where the cross section $\sigma$ is assumed to be the geometric one $\sigma=4\pi R_e^2$.  A convenient way to estimate $t_h$ is to consider the mean free path
for a photon before colliding with a baryon. This is given by \cite{kolb}
$$
t_b=\frac{1}{x_e n_B\sigma_T}\sim 3.9 \;10^{18}\frac{a^3}{\Omega_Bh^2}s,
$$
with $x_e$ the fraction of ionised particles and $\sigma_T$ the Thompson cross section for baryons.
Thus
$$
t_h\sim\frac{\sigma_T}{4\pi R^2}\frac{n_B}{n_{\text{dm}}} 3.9 \;10^{18}\frac{a^3}{\Omega_Bh^2}s.
$$
This can be expressed in terms of $\Omega_{dm}h^2$ by use of (\ref{magic}) as follows
$$
t_h\sim\frac{\sigma_T}{4\pi R^2} 3.9 \;10^{18}\frac{a^3}{\Omega_{\text{dm}}h^2}s\frac{m_{\text{dm}}}{m_N}.
$$
On the other hand, at the time of matter radiation equality
$$
a_{eq}\sim \frac{4.15\;10^{-5}}{\Omega_m h^2}\sim 3.5\;10^{-4},\qquad T_{eq}\sim 5.7 \Omega_m h^2 \text{eV}\sim 0.73 \text{eV},
$$
where the estimation $\Omega_m h^2\sim 0.128$ has been employed. As $a\sim T^{-1}$ it follows that
$$
a\sim \frac{2.55\;\text{eV}}{T}10^{-4}.
$$
By taking into account that $\sigma_T\sim 2.4\pi m_N^{-2}$, that $R_e\sim 10^{-5}$ cm and that $m_{\text{dm}}\sim m_N B$, it follows 
that 
$$
t_h\sim \frac{10^{17}}{\Omega_{\text{dm}}h^2}\bigg(\frac{2.55\text{eV}}{T}\bigg)^3 s.
$$
Here the value $B\sim 10^{29}$ was employed, which corresponds to $R_e\sim 10^{-5}$ cm.
This time is larger than the Hubble time
$$
H^{-1}\sim 1.13\; 10^{12}\bigg(\frac{\text{eV}}{T}\bigg)^{\frac{3}{2}} \frac{s}{\sqrt{\Omega h^2}},
$$
thus the axion quark nuggets may be considered components of cold dark matter even when they strongly interact with light. In other words, it takes really a long time
for a photon for reaching the compact object. 
\\

\emph{Energy injection of axion quark nugget}
\\

Consider the universe a temperatures of the order $T>2 m_e\sim$ MeV, which hold at ages right before Big Bang Nucleosynthesis. At this stage, the universe is composed by a plasma of electrons, positrons, photons and baryons. Some electrons, whose density is given by $n_e$, will be annihilated due to the large number of positron on the electrosphere of the nugget \cite{zhitnitsky12}. The microscopic description of this electrosphere may be found in the works \cite{zhitnitsky12}, \cite{electrosphere}.  These annihilations would lead to an energy injection in the plasma. It is important that this injection remains small, otherwise the presence of the nugget may alter considerably the pre Big Bang Nucleosynthesis cosmology.

The number of events by unit time between a given nugget and the plasma can be estimated by the following formula
$$
\frac{dN}{dt}\sim 4\pi R^2 n_e.
$$
A typical  electron or positron energy in this plasma is of the order $\mu_{e^+}\sim 10$ MeV. By taking into account the last formula, it follows that the energy injection to the plasma by unit volume due to nuggets is
$$
\frac{dE}{dV dt}\sim 4\pi R^2 n_e \mu_{e^+} n_{qn}.
$$
This annihilation will result in an event with typical energy $\mu_{e^+}$. On the other hand, a typical energy energy density of the system $T n_e$ and a typical time between collisions is $\tau\sim \alpha^{-2} T^{-1}$, with $\alpha$ the standard QED coupling.
The dimensionless quotient
$$
\frac{\tau}{T n_e}\frac{dE}{dV dt}\sim \frac{4\pi \tau R^2 \mu_e n_{qn}}{T},
$$
compares the energy injected by annihilations with a typical energy in the plasma. The density of nuggets, as stated above, is approximately
$$
n_{qn}\sim \frac{n_B}{\overline{B}}.
$$
The average nugget baryon charge is, as stated in the text, is $\overline{B}\sim 10^{28}$. The baryon number $n_B$, in the standard cosmology context, is
given by
$$
n_B\sim \eta \;n_\gamma, \qquad \eta\sim 5.10^{-10},\qquad n_\gamma\sim n_e\sim n_{e^+} \sim \frac{2}{\pi^2} T^3.
$$
This means that
$$
\frac{\tau}{T n_e}\frac{dE}{d.V dt}\sim \frac{8 R_n^2 \mu_e \eta T}{\pi \alpha^2\overline{B}}.
$$
For the parameters found in the present work at $T\sim m_e$
$$
\frac{\tau}{T n_e}\frac{dE}{d.V dt}\sim 10^{-22}.
$$
The energy injection is therefore completely subdominant with respect of a typical energy of the plasma, which is the desired result.
 \\

\emph{A comment about the Lithium puzzle}
\\

At temperatures high enough, a given nugget may become ionized by striping off positrons of its electrosphere. Some protons present in the universe may then be influenced by the nugget  electric field and will attempt to screen its charge.The same holds for ions with $Z>1$.  The presence of nuggets should not modify considerably the  proton density of the universe.  On the other hand, if the density of  ion species with $Z>1$ is depleted, this may have interesting consequences in the context of the lithium puzzle \cite{zhitnitsky12}. But even before to discuss these matters, it is important to remark that \cite{zhitnitsky12} employs a nugget radius
of the order $R_n\sim 10^{-5}$ cm while the present one uses $R_n\sim 10^{-6}$ cm, that is, one order of magnitude less. Our claim is that, even taking into account this discrepancy, there is room to assess that the proton density remains unchanged and that the lithium density is depleted. The main point is the definition of the so called capture radius, which it is described below.

The microscopic profile of the nugget electrosphere is described detail in \cite{electrosphere}. If the nugget is placed at temperatures high enough $T>m_e$, the positrons with momentum $p^2_+<2m_{e}T$ are expected to be stripped off, since they are weakly bound to the object.  The nugget becomes ionized, and the charge $Q(r)$ enclosed in a sphere radius $r$ is characterized in \cite{zhitnitsky12}.  For temperatures  $T<100$ MeV, the more massive protons which will attempt to screen the nugget charge $Q(r)$. In fact, the protons are influenced under the nugget electric potential energy, and those which are close enough to the object will have negative energy, thus bounded to the object.
It is clear that  other charged particles with $Z>1$ may be trapped as well. 

Consider a nucleus specie with  $Z>1$. The density variation of such specie in presence of nuggets is estimated in \cite{zhitnitsky12} as
\be\lb{copo}
\frac{\delta n_Z}{n_Z}\sim \frac{4\pi R^3_c(T) n_{qn}}{3} e^{\frac{(Z-1)\alpha Q(R_c)}{R_c T}}.
\ee
The first factor is the density variation of protons, whose $Z=1$, and is proportional to the nugget density $n_{qn}$ and to the volume of the sphere enclosed by the so called capture radius $R_c$.  The exponential factor represents the enhancement of the density variation due to the larger Coulomb interaction of nucleus with $Z>1$ with respect to the protons. 
The correct definition of the capture radius $R_c(T)$  is subtle. A possibility is to define a radius such that for $r<R_c(T)$ the electrostatic energy for protons is such that $2 Q(r)>m^2 r v\sim 2Tr$. Clearly, in this case, the proton energy will be negative and this particle will be bounded to the object. Another possible definition is that for $r>R_c(T)$ the density of protons $n_p(r, T)$ approaches the cosmological value $n_B(T)$. The capture radius may be approximated by the following condition
\be\lb{terr}
n_p(R_c, T)\sim n_B(T)\sim \frac{2\eta}{\pi^2} T^3.
\ee
This condition means that,  for $R$ larger than the capture radius $R_c$, the proton density is approximately equal to the cosmological baryon number $n_B(T)$.
In other words, this density is only modified for regions closer to the nugget. On the other hand, the radius dependence of the proton density may be estimated as
$$
n_p(R, T)\sim n_p(R_n, T)\bigg(\frac{R_n}{R}\bigg)^p,
$$
with $p$ an unknown exponent which, in reference \cite{zhitnitsky12}, is approximated by $p\sim 6$.  The density of protons $n_p(R_n, T)$ at the radius nuggets is very high, and can be estimated
by effective approximation schemes such as Thomas-Fermi method \cite{zhitnitsky12}, \cite{electrosphere}. The density $n_p(R_n, T)$ is then fixed by the condition that the protons screen a large portion of this charge. 
The result \cite{zhitnitsky12}
$$
n_p(R_n, T)\sim \frac{m_e T^2}{\pi \alpha},
$$
does not depend on the nugget size $R_n$. Besides, the charge of the ionized nugget is \cite{zhitnitsky12}
$$
Q_i\sim \frac{2\sqrt{2}R_n^2}{\alpha} T^{\frac{3}{2}}\sqrt{m_e}.
$$
From the parameters found in the present work, one has from (\ref{terr}), that $R_c\sim 10^{-4}$ cm  at temperatures relevant for lithium physics $T\sim 20$ KeV. 
We suggest however, that the  radius that should be employed in (\ref{copo}) may be slightly larger. First, the functional form for $n_p(R, T)$ as a $p$-power is a good approximation, but it is expected to be corrected at some distance from the nugget. On the other hand, the ionized charge that is screened by the proton at the capture radius is \cite{zhitnitsky12}
$$
Q(R_c)\sim \int_{R_c}^\infty n_p(R, T)4\pi r^2 dr\sim \frac{4\pi n_B(T)R_c^3}{p-3}.
$$
With the numbers found here one has that $Q_i\sim 10^9$ and $Q(R_c)\sim 10^{6}$, that is, three orders smaller. It seems reasonable for the authors to employ a radius in (\ref{copo}) for which the charge is more screened than that. The three orders of magnitude discrepancy may be corrected by employing a radius close to $R'_c\sim 10^{-3}$ cm. Here the dot is emphasizing that we are not necessarily identifying this radius with the capture one. The value $10^{-3}$cm  is very close to the numerical value that \cite{zhitnitsky12} employs. By parameterizing the $Z=1$ density as
$$
\frac{4\pi R'^3_c n_{qn}}{3}\sim e^{-X_{p}},\qquad X_{p}=-\log\frac{4\pi R'^3_c n_{qn}}{3},
$$
and by introducing the enhancement exponent
$$
X_e=\frac{(Z-1)\alpha Q(R'_c)}{R'_c T},
$$
it is found, by assuming the average value $B\sim 10^{28}$, that
$$
X_p\sim 39.5-\bigg(3-\frac{3}{p}\bigg)\log\frac{T}{20\text{KeV}},\qquad X_e\sim 20 (Z-1)\bigg(\frac{T}{20\text{KeV}}\bigg)^{2(1-\frac{1}{p})}.
$$
At temperatures $T\sim 20$KeV, the factor $X_p\sim 39.5$ and thus, $\delta n_p/n_p\sim 10^{-39.5}$. Thus the nugget do not alter the proton density, which is a desired feature.
On the other hand, for $Z=3$ it follows that both factors $X_p$ and $X_e$ are roughly the same at $T\sim 20$KeV, and both close to the value $40$.  Even more, for $Z>3$ the enhancement factor $X_e$ becomes dominant. This suggests that for species with $Z\geq 3$ the deviation $\delta n_Z\geq n_Z$. Thus, for such charged ions, it is expected that the density $n_Z$ is considerably depleted. Therefore the applications to the lithium puzzle discussed in \cite{zhitnitsky12} does not seem to be spoiled by the nugget description presented in this work. We hope to come with a more detailed analysis of this issue in a future work. Another point that deserves to be studied further is the bias between nuggets and anti-nuggets. 
The results of \cite{zhitnitsky6} suggest that axionic field variations
of small scale are possible at QCD scale. In authors opinion, this possibility is 
attractive since is suggesting that the bias is due to CP violating physics, and this conjecture seems natural. But clearly to put these ideas in more precise quantitative form is of interest for a future. 
\\

\emph{Some alternatives to the present model}
\\

It should be emphasized that the original model \cite{zhitnitsky}-\cite{zhitnitsky12} is based on closed bubbles with no string attached, which appear naturally in the Kibble model. As argued along the text, this leads to a reasonable density of nuggets and to interesting phenomenological consequences. However, it is of interest to describe, at least qualitatively, the physics that is obtained from other initial scenarios such as the ones corresponding to domain walls with strings attached, which do not necessarily rely on the Kibble mechanism, as for instance \cite{sikivie}.  In the standard picture for this objects \cite{sikivie} the unique
genuinely topological object in this model (with N=1 axion model) is the axion string. These defects are originated
at the PQ transition, which corresponds to a very large temperature. Axionic domain
wall arises near QCD scale when QCD instanton effects tilt the axion
potential. A large number of these domain walls happen to be open, and axion strings
become their boundary. This implies that the number and size of
these axion domain walls at QCD scale will be entirely determined by
the number of axion strings at that scale, which would have entered
the scaling solution by that time. This will lead to typically of
order 10 axionic strings, hence similar number of axionic domain walls
in the horizon volume at QCD scale \cite{sikivie}. 

The density described in the previous paragraph implies that the closed domains wall that are initially present are typically of a radius $R_0\sim 10^2$ meters.
We have considered the evolution of a large initial bubble, with initial radius of $R_0\sim 10^2$ meters. Without quoting all the details, it should be said that initially it is the neutrino emissivity that dominates the dynamics, until the radius is of the order $R_0\sim 10^{-2}$ meters. At this stages $\mu \sim T$. After that, the object stabilizes at a radius of the order
$R_e\sim 10^{-2}-10^{-3}$ cm if the axion coupling constant is chosen of the order $f_a\sim 10^{9}-10^{10}$ GeV. The value of the resulting time scale $\tau$ that results from (\ref{nomers}) is essentially the same as for a small bubble namely, $\tau\sim 10^{-4}$ cm. In other words, we suggest that  if the initial radius $R_0$ is varied from $100$ meters to $1$ cm them, by varying
the axion coupling in the range $10^{9}$ GeV$<f_a<10^{12}$ GeV, the desired value of $\tau$ may be obtained, even for this large initial radius.

The fact that the evolution of the nuggets described above can give raise to an appropriate value of $\tau$ as well is of interest. However, another stern test about the physics of these objects
is their density. There are indications that \cite{sikivie2}-\cite{sikivie3} that only a very small fraction of the resulting domain walls are closed. A characteristic fraction
may be $10^{-7}$. While it is not impossible that this number may be enough for dark matter generation, it may be of interest to 
consider other possibilities, in which the number of defects is enhanced. There appeared recently  literature \cite{manohar} concerning these matters and, in particular, to the enhancement of the number of string per Hubble horizon. The reference \cite{manohar} states, by use of anomaly arguments of the Callan-Harvey type \cite{harvey}, that axion strings are superconducting. The Callan-Harvey anomalies in these defects are responsible, in particular, of effects such as current leakage.  A shrinking axion loop evolves into a vorton, whose stability is supported by the electromagnetic force on the string current. If there is a primordial magnetic field at the stages of string formation, a large current is induced
on the axion string, and there appears a further drag force with the surrounding particles. As a consequence, the string movement is slowed down. Depending on the value of the primordial magnetic field, this may give rise to a large enhancement of the number strings per horizon. Another source of enhancement is likely to appear at the PQ phase transition.
The axion loops become the boundary of a domain wall. The shrink of the loop is stopped at some point by  the electromagnetic force generated by the current. However, the wall is still shrinking and the system is twisting violently, which leads to a breaking into smaller pieces of vortons. This process keeps going till the size of the vorton becomes arguably less than the axion Compton wavelength. This implies an enhancement of the number of these objects  It is suggested in \cite{manohar} that the resulting domain wall dynamics is not affected considerably by the presence of these magnetic fields. Thus, these defects seems to be promising if there are primordial magnetic fields present at the hadron-quark phase transition, as their density is enhanced. This may lead to an interesting line for future investigation. 

The last thing to discuss in these alternative scenarios is the excess of nuggets over anti nuggets. As in the previous case, a further elaboration of the picture is needed.
To introduce such excess, coherence of axion field
over the entire universe is required. A possibility is to assume that PQ symmetry breaking occurs before inflation. However, that
will also wash out any axionic strings, hence associated axionic
domain walls. Nevertheless, this may not be the case if there is a proper enhancement such as the one induced by magnetic field. In this case there appears a competition
between the expansion of the universe and the increase of objects by the magnetic field.
Axionic domain walls only arise at QCD scale by
concentrating the angular variation around the string in a wall
(actually a wedge). But a more firm quantitative analysis is desirable before believing in such conclusions, in particular related to this competition of effects. We hope to come with a better
analysis in a future investigation.

\section*{Acknowledgments}
O. S is supported by CONICET, Argentina.

\end{document}